\documentclass{amsart}

\newtheorem{thm}{Theorem}[section]
\newtheorem{cor}[thm]{Corollary}

\theoremstyle{definition}
\newtheorem{defn}[thm]{Definition}
\theoremstyle{remark}

\numberwithin{equation}{section}

\newcommand{\N}{\mathbb{N}}
\newcommand{\Q}{\mathbb{Q}}
\newcommand{\R}{\mathbb{R}}

\newcommand{\abs}[1]{\left\vert#1\right\vert}

\newcommand{\seq}[1]{\left\{#1\right\}}

\begin{document}
\include{siamltex}
\title{Using biased coins as oracles}
\author{Toby Ord}
\address{Faculty of Philosophy, University of Oxford, Oxford, OX1 4JJ, UK}
\email{toby.ord@philosophy.ox.ac.uk}
\author{Tien D. Kieu}
\address{Centre for Atom Optics and Ultrafast Spectroscopy, Swinburne University 
of Technology,  Hawthorn 3122, Australia}
\email{kieu@swin.edu.au}


\begin{abstract}
While it is well known that a Turing machine equipped with the ability to flip a 
fair coin cannot compute more that a standard Turing machine, we show that this 
is not true for a biased coin. Indeed, any oracle set $X$ may be coded as a 
probability $p_{X}$ such that if a Turing machine is given a coin which lands 
heads with probability $p_{X}$ it can compute any function recursive in $X$ with 
arbitrarily high probability. We also show how the assumption of a non-recursive 
bias can be weakened by using a sequence of increasingly accurate recursive 
biases or by choosing the bias at random from a distribution with a 
non-recursive mean. We conclude by briefly mentioning some implications 
regarding the physical realisability of such methods.
\end{abstract}

\maketitle
\begin{center}
\today
\end{center}





\section{Introduction and Motivation}

The Turing machine is well known to be a very robust model of computation. In 
almost all textbooks on the theory of computation, one can find a list of 
extensions to the Turing machine that offer it more primitive resources, such as 
extra tapes or nondeterminism, and yet do not give it the ability to compute any 
additional functions. Amongst such resources it is not uncommon to find 
references to probabilistic methods such as coin tossing. 

These methods can be made precise with the introduction of the probabilistic 
Turing machine or PTM \cite{Gill:1977}. A PTM is a standard Turing machine with 
a special randomising state. When the machine is in this state, the transition 
to a new state is not governed by what is on the tape, but by a random event. A 
fair coin is tossed and the machine goes to the specified 1-state if the coin 
comes up heads and the 0-state if it comes up tails.

Unlike a Turing machine, a PTM will not necessarily return the same output when 
run multiple times on the same input. Care must therefore be taken in defining 
what it means for a function to be computed by a PTM. One way is to say that a 
PTM computes a given function, $f$, is that if given $x$ as input, along with a 
measure of accuracy $j \in \N$, it produces $f(x)$ with probability at least 
$1-\frac{1}{2^{j}}$. By this definition, a function is computable by a PTM if 
and only if it can be computed with arbitrarily high confidence. Alternatively, 
we could relax this definition and say that a PTM computes $f$ if and only if 
when given $x$ as input, it produces $f(x)$ with some probability greater than 
$\frac{1}{2}$.

It is quite easy to see that with either definition, a PTM computes only the 
recursive functions. For any PTM $P$, there is a Turing machine $T$ that 
simulates it. $T$ simulates each branch of the computation in parallel and keeps 
track of their respective probabilities. $T$ also keeps a table which associates 
outputs with their probabilities. When a branch halts returning some value $y$, 
$T$ creates a new position in the table for $y$ and stores the probability of 
that branch occurring. If a branch has already halted with output $y$, $T$ simply
adds the new probability of producing $y$ to the old value. After each update to 
the table, $T$ checks whether the new value for $y$ is greater than $P$'s 
threshold ($\frac{1}{2}$ or $1-\frac{1}{2^{j}}$) and halts returning $y$ if this 
is so. In this way, $T$ halts with output $y$ if and only if $P$ returns $y$ 
with sufficient probability.

This argument can also be extended to deal with more complicated probabilistic 
methods. For example, we could allow biased coins where the chance that heads 
comes up is some given rational number. We could even allow the bias to be any 
recursive real number (as defined in section \ref{Section:ApproximatingP} of 
this paper). In each case, $T$ can still keep track of the probability of each 
computation branch and test to see whether an output occurs with enough 
probability to be deemed $the$ output of the PTM.

It is important to ask, however, what can be computed if non-recursive 
probabilities are used. In this paper, we show that allowing coins with 
non-recursive biases makes the above argument fail quite spectacularly. We first 
show that a PTM can compute arbitrarily accurate estimates to the bias on its 
coin and then strengthen this to computing arbitrarily many bits of the binary 
expansion of the bias.\footnote{Since writing this paper, an article by Santos 
\cite{Santos:1971} has been brought to the authors' attention wherein a similar 
result is shown. However, it is our opinion that Santos' proof is incomplete, 
lacking an explanation of how the binary expansion of the probability can be 
computed from the rational approximations. In any event, we think the present 
account is useful for its further results and use of only elementary methods.} 
From this, we reach several strong theorems about the power of PTMs, showing in 
particular that there is a single PTM that acts as a universal $o$-machine: when 
equipped with a probability coding a given oracle, it simulates a given 
$o$-machine with that oracle on a given input to a given level of confidence. 
Thus, the addition of randomness to the resources of a Turing machine can 
certainly increase its computable functions. Only when the coins are restricted 
to recursive biases does it offer no additional power.

In the remaining sections, we show two ways in which the same results are 
possible with slightly weakened resources. Specifically, we show how a sequence 
of rationally biased coins can be used, so long as the biases converge 
effectively to a non-recursive real or the biases are drawn at random from a 
distribution with a non-recursive mean. Finally, we point to some interesting 
physical applications in which these types of probabilistic methods seem to be 
consistent with Quantum mechanics.


\section{Approximating $p$ to arbitrary accuracy}
\label{Section:ApproximatingP}

The natural way to approximate the probability $p$ that the coin will land 
heads, is to look at the average number of heads in $n$ tosses. By the weak law 
of large numbers, this value  (which we will denote $\hat{p}$) approaches $p$ as 
$n$ approaches infinity. However, to approximate $p$ effectively, we need to 
know how fast this convergence is likely to be. This can be expressed by asking 
how many tosses are required before $\hat{p}$ is within a given distance of $p$ 
with a given level of confidence. Specifically, we will ask for a method of 
calculating $n$ such that when at least $n$ tosses are made, $\abs{\hat{p} - p} 
< \frac{1}{2^{k}}$ with probability at least $1-\frac{1}{2^{j}}$ for given $j,k 
\in \N$.

The probability distribution of possible values of $\hat{p}$ for a given value 
of $n$ is a binomial distribution with mean $p$. The variance of $\hat{p}$ is 
given by
\begin{equation}
\sigma^{2}
=\frac{p(1-p)}{n}
\label{BinomialVariance}
\end{equation}

This variance depends upon the unknown value of $p$, however since it has a 
maximum where $p=\frac{1}{2}$, we can see that
\begin{equation}
\sigma^{2} \le \frac{1}{4n}
\label{BinomialVarianceBound}
\end{equation}

With this upper bound for the variance, we can use the Chebyshev inequality
\begin{equation}
\forall \epsilon \ge 0\quad P(\abs{x-\mu} \ge \epsilon) \le
\frac{\sigma^{2}}{\epsilon^{2}}
\label{ChebyshevInequaltiy}
\end{equation}
to form an upper bound for the probability of error
\begin{equation}
\forall k \quad P\left(\abs{\hat{p}-p} \ge \frac{1}{2^{k}}\right) \le
\frac{2^{2k}}{4n}
\label{VarianceInequaltiy}
\end{equation}

Therefore, if we insist on a chance of error of at most $\frac{1}{2^{j}}$, this 
can be achieved so long as
\begin{eqnarray}
\frac{2^{2k}}{4n} & \le & \frac{1}{2^{j}}\\
n & \ge & 2^{j+2k-2}
\label{nConstraint}
\end{eqnarray}

Thus, for each value of $p \in [0,1]$ we can compute an approximation of $p$ 
that is within an arbitrarily small distance of the true value with an 
arbitrarily high probability. More formally,

\begin{thm}
\label{CauchyPTM1}
There is a specific PTM that, when equipped with a probability $p$, takes inputs 
$j,k \in \N$ and outputs a rational approximation to $p$ that is within 
$\frac{1}{2^{k}}$ of the true value with probability at least 
$1-\frac{1}{2^{j}}$.
\end{thm}
\begin{proof}
The PTM simply tosses its coin $2^{j+2k-2}$ times and returns the ratio of heads 
to tails. By the argument above, this approximation will suffice.
\end{proof}
 
This method of approximating a real number by successively accurate rational 
approximations can also be used to define a notion of which real numbers are 
computable by a (deterministic) Turing machine. For convenience, we say
\begin{defn}
$\seq{x_{n}}$ \emph{converges quickly} to $x$ if and only if 
$\abs{x_{n}-x}<\frac{1}{2^{n}}$ for all $n$.
\end{defn}

We can then define a notion of a recursive real.
\begin{defn}
$x \in \R$ is recursive if and only if there is a Turing machine that takes $n 
\in \N$ as input and returns $x_{n} \in \Q$, where $\seq{x_{n}}$ converges 
quickly to $x$.
\end{defn}

The recursive reals given by this definition are well studied and include a 
great many of the reals actually encountered in mathematics, including all the 
algebraic numbers as well as $\pi$ and $e$. However, since there are uncountably 
many reals but only countably many Turing machines, it is clear that most of 
them are not recursive. If a PTM is equipped with one of these non-recursive 
reals as its probability, then our algorithm of the previous section shows that 
in a certain sense, this PTM can compute this real --- a feat that is impossible 
with a deterministic Turing machine.

However, there is still some room to question whether the PTM of Theorem 
\ref{CauchyPTM1} actually computes its probability. Consider, for example, the 
following alternative definition of a recursive real.
\begin{defn}
$x \in \R$ is recursive if and only if there is a Turing machine that takes no 
input and outputs a sequence $\seq{x_{n}}$ which converges quickly to $x$.
\end{defn}

This definition is evidently equivalent to the previous one when it comes to 
deterministic Turing machines, but it is not immediately clear that the 
equivalence holds for PTMs. While the PTM of Theorem \ref{CauchyPTM1} can 
compute each approximation to $x$ with arbitrary accuracy, it is not clear that 
a PTM could output an infinite sequence of approximations with them \emph{all} 
being correct with arbitrarily high probability. However, we now show that this 
can be achieved by requiring each successive event to be more and more probable.

For a given minimum probability $q$ for an entire infinite sequence of events 
occurring, we can set the probability of the $i$-th event occurring $q_{i} = 
q^{2^{-i}}$. It follows that the chance of all events occurring is
\begin{equation} 
\prod_{i = 1}^{\infty}q^{2^{-i}} = q^{\sum_{i=1}^{\infty}2^{-i}} = q
\label{InfiniteProduct}
\end{equation}

In addition, we can consider the chance that all events in an infinite suffix of 
the sequence occur. The chance of all events after event $N$ occurring is
\begin{equation} 
\prod_{i = N+1}^{\infty}q^{2^{-i}} = q^{\sum_{i=N+1}^{\infty}2^{-i}} = 
q^{2^{-N}}
\label{InfiniteSuffixProduct}
\end{equation}
Thus, for each $\epsilon > 0$, there is a value of $N$ for which the probability 
of all events after event $N$ occurring is within $\epsilon$ of 1. Therefore, the 
probability that some infinite suffix of these events will occur must be equal 
to 1.

This construction can be applied in the case of our approximations to $p$. In 
particular, we can find a new value $j'$ as a function of $j$ and $k$ which can 
then be substituted into our formula for the number of required coin tosses.
\begin{equation} 
1-\frac{1}{2^{j'}} = \left(1-\frac{1}{2^{j}}\right)^{2^{-k}}
\label{NewJValue1}
\end{equation}

Using a Taylor expansion, we can see that for $0 < x,y < 1$
\begin{equation} 
(1 - x)^{y} < 1 - xy 
\label{HelpfulEquation}
\end{equation}
and thus
\begin{eqnarray} 
1-\frac{1}{2^{j'}} & < & 1-\left(\frac{1}{2^{j}}\right)
\left(\frac{1}{2^{k}}\right)\\
1-\frac{1}{2^{j'}} & < & 1-\frac{1}{2^{j+k}}\\
j' & < & j+k
\label{NewJValue2}
\end{eqnarray}

Putting this all together,
\begin{thm}
There is a specific PTM that, when equipped with a probability $p$, takes input 
$j \in \N$ and outputs a sequence $\seq{\hat{p}_{k}}$ that converges quickly to 
$p$. Furthermore, with probability 1, there is some $N$ such that 
$\abs{\hat{p}_{k} - p} < \frac{1}{2^{k}}$ for all $k > N$.
\end{thm}
\begin{proof}
For each value in the sequence, the PTM simply tosses its coin $2^{j+3k-2}$ 
times and returns the ratio of heads to tails. By the argument above, these 
approximations will suffice.
\end{proof}


\section{Computing the binary expansion of $p$}
\label{Section:BinaryExpansion}

The definitions of the previous section are not the only ways that the recursive 
reals can be defined. Instead of using converging sequences of rationals, we can 
use the original technique due to Turing \cite{Turing:1936} of using the base 
$b$ expansion. For simplicity, we use the binary expansion and only consider 
those reals in the unit interval.

\begin{defn}
$x \in \R$ is recursive if and only if there is a Turing machine that takes $n 
\in \N$ as input and returns $b_{n}$, the $n$-th bit of the binary expansion of 
$x$.
\end{defn}

As before, we can rephrase this to speak of Turing machines that take no input:

\begin{defn}
$x \in \R$ is recursive if and only if there is a Turing machine that takes no 
input and returns the sequence $\seq{b_{n}}$, coresponding to the binary 
expansion of $x$.
\end{defn}

Both definitions run into an ambiguity in the case of \emph{dyadic} rationals: 
those that can be expressed in the form $\frac{n}{2^{m}}$. For such numbers, 
there are two binary expansions so we adopt the convention of using the one 
containing an infinite number of 0's.

By extending our method for approximating $p$, we can also approximate the 
binary expansion of $p$. Unfortunately this will not be possible if $p$ is a 
dyadic rational, so for now consider the case where it is not, and $p$ thus has 
a unique infinite binary expansion in which both 0 and 1 occur infinitely many 
times. To compute the binary expansion of $p$, we need a method that takes 
inputs $j,l$ and gives us the value of $b_{l}$ with probability 
$1-\frac{1}{2^{j}}$.

It may seem as though this can be achieved simply by computing $\hat{p}_{l+1}$ 
and taking its $l$-th bit, but problems arise when a run of consecutive 0's or 
1's occurs around this point in the expansion. For instance, if we want the 
third bit and $\hat{p}_{4} = .01111111$, then the true value of $p$ could be as 
low as $.01101111$ or as high as $.10001111$ and we can thus be certain of none 
of the bits. By using the following algorithm, which we shall call $A$, we can 
overcome this problem.\\

\begin{itemize}
\item $k := l$
\item repeat
\begin{itemize}
\item $k := k+1$
\item compute $\hat{p}_{k}$ (by tossing the coin $2^{j+3k-2}$ times)
\item if $\hat{p}_{k}<1$ and there are both a 0 and a 1 between the $l$-th and 
$k$-th bits of the expansion of $\hat{p}_{k}$ then output the $l$-th bit\\

\end{itemize}
\end{itemize}

An analysis of $A$ is made somewhat complex by the fact that it involves random 
events and does not always give the correct output, but for now we will just 
consider the most probable case where the probabilistically generated sequence 
$\seq{\hat{p}_{k}}$ converges quickly to $p$. We can see that there must be a 
value of $k$ for which $\hat{p}_{k}$ is less than one and has both a 0 and a 1 
between its $l$-th and $k$-th bits, for if there were not then $\hat{p}_{k}$ 
would either be approaching a dyadic rational or failing to converge --- each of 
which would contradict our assumptions. Therefore, so long as $p$ is not a 
dyadic rational and $\seq{\hat{p}_{k}}$ converges quickly to $p$, $B$ will 
always halt. When it does, the value of $\hat{p}_{k}$ will be in the form
\begin{equation} 
\hat{p}_{k} = .b_{1} \ldots b_{l} 1 \ldots 1 0 b_{k} \ldots
\label{BinaryFormOne}
\end{equation}
or
\begin{equation} 
\hat{p}_{k} = .b_{1} \ldots b_{l} 0 \ldots 0 1 b_{k} \ldots
\label{BinaryFormTwo}
\end{equation}

In either case, adding or subtracting a value smaller than $\frac{1}{2^{k}}$ 
will not change any of the first $l$ bits of $\hat{p}_{k}$ and since $p$ is 
within $\frac{1}{2^{k}}$ of $\hat{p}_{k}$, their first $l$ bits must be 
identical.

It is important to note, however, that while all runs of 1's or 0's within the 
expansion of $p$ must come to an end, they can be arbitrarily long, so the 
running time of $A$ depends upon the value of $p$. If the $l$-th bit of the 
expansion of $p$ is followed by a run of $m$ identical bits, then we must 
compute $l+m$ values of $\hat{p}_{k}$, requiring at most $2^{j+3l+3m-1}$ coin 
tosses.

What about those cases where $\seq{\hat{p}_{k}}$ does not converge quickly to 
$p$? This can be for two different reasons --- either it converges to $p$, but 
not as quickly as required or it does not converge to $p$ at all. The first of 
these cases occurs with probability $\frac{1}{2^{j}}$ and while it cannot cause 
$B$ to fail to halt, it may well cause an incorrect output. The second case 
occurs only with probability 0, and may either cause an incorrect output or 
non-termination.

\begin{thm}
There is a PTM that implements $A$. Equipped with any non-dyadic probability 
$p$, it takes positive integers $j$ and $l$, outputting the $l$-th bit of the 
binary expansion of $p$ with probability greater than $1-\frac{1}{2^{j}}$. The 
probability that it returns an incorrect answer is less than $\frac{1}{2^{j}}$, 
while the probability that it does not terminate is 0.
\end{thm}
\begin{proof}
Immediate
\end{proof}

We can also modify $A$ to form $A_{\infty}$ which takes only $j$ as input and 
outputs the entire expansion of $p$. In this case it outputs the $l$-th digit 
when it has output all prior digits and has found a value of $\hat{p}_{k}$ with 
a 0 and a 1 between its $l$-th and $k$-th digits. $A_{\infty}$ uses the high 
likelyhood of $\seq{\hat{p}_{k}}$ converging quickly to $p$ to greater effect 
than $A$, by generating the \emph{entire} expansion with arbitrarily high 
probability

\begin{thm}
There is a PTM that implements $A_{\infty}$. Equipped with any non-dyadic 
probability $p$, it takes a positive integer $j$, outputting the entire binary 
expansion of $p$ with probability greater than $1-\frac{1}{2^{j}}$. The 
probability that it outputs finitely many incorrect bits is less than 
$\frac{1}{2^{j}}$, while the probability that it outputs infinitely many 
incorrect bits or outputs only a finite number of bits is 0.
\end{thm}
\begin{proof}
Immediate
\end{proof}


\section{Using the binary expansion of $p$ as an oracle}
\label{Section:PAsOracle}

In 1939, Alan Turing \cite{Turing:1939} introduced a very influential extension 
to his theoretical computing machines. Turing's $o$-machines are standard Turing 
machines combined with a special `oracle', which can answer questions about a 
particular set of natural numbers, called its oracle set. Like a PTM, an 
$o$-machine has a special query state and two answer states, but instead of the 
answer being given randomly, it corresponds to whether a certain number is in 
the oracle set. To specify the number whose membership is being questioned, a 
special symbol $\mu$ is inscribed twice on the tape and the number of squares 
between each inscription of $\mu$ is taken as the query to the oracle. Depending 
on which oracle set is given, an $o$-machine can compute different classes of 
functions, and they thus give rise to a notion of relative computability.

Corresponding to an $o$-machine with oracle $X$ we can construct a PTM with 
probability $p_{X}$ where the $n$-th digit of the binary expansion of $p_{X}$ is 
1 if $n \in X$ and 0 otherwise. A PTM equipped with $p_{X}$ can perform all 
basic operations of a Turing machine, as well as determining whether $n \in X$ 
for any $n$. It can do this by simulating $B_{\infty}$ in parallel with its main 
computation, storing the bits of $p$ produced by $B_{\infty}$ and examining them 
when needed. If it needs to test whether $n \in X$ and has not yet determined 
$b_{n}$, it simply waits until this is found.

In the cases where $p_{X}$ is a dyadic rational this method will not work, but 
since $X$ will be recursive, there is a probabilistic Turing machine that can 
simulate such an $o$-machine without using any probabilistic methods at all. In 
this way, these methods suffice to simulate any $o$-machine.

\begin{thm}
For any $o$-machine $M$ with oracle $X$, there is a PTM $P_{M}$ equipped with 
probability $p_{X}$ that when given the same inputs plus one additional input 
$j$, $P_{M}$ produces the same output as $M$ with probability greater than 
$1-\frac{1}{2^{j}}$.
\end{thm}
\begin{proof}
Immediate
\end{proof}

Since all functions of the form $f:\N^{n} \to \N^{m}$ or $f:\N^{n} \to \R^{m}$ 
are computable by some $o$-machine, we can see that there are probabilities that 
would allow PTMs to compute any such functions.

\begin{cor}
For any function $f:\N^{n} \to \N^{m}$ or $f:\N{n} \to \R^{m}$, there exists a 
PTM that when given inputs $j,x_{1},\ldots,x_{n}$ produces 
$f(x_{1},\ldots,x_{n})$ with probability greater than $1-\frac{1}{2^{j}}$, 
produces incorrect output with probability less than $\frac{1}{2^{j}}$ and 
diverges with probability 0.
\end{cor}

Since these natural and real numbers can be used to code other mathematical 
objects, this set of PTM computable functions includes a vast number of 
interesting mathematical functions. Given an appropriately biased coin, a PTM 
could decide the halting problem or the truths of first order arithmetic.

Finally, just as there is a single universal Turing machine which can take the 
code of a Turing machine as input and simulate it, so there is a universal 
$o$-machine which takes the code of an arbitrary $o$-machine and simulates it 
\emph{so long as it is equipped with the oracle of the machine being simulated}. 
A similar job can be performed by a specific PTM, provided that the $o$-machine 
to be simulated does not have an oracle set that would be encoded as a dyadic 
rational. As such $o$-machines can only compute recursive functions, this is not 
a great concern.

\begin{thm}
There is a specific PTM $P_{U}$ that takes inputs $j,n,m \in \N$ and when 
equipped with any non-dyadic probability $p_{X}$, $P_{U}$ computes the result of 
applying the $o$-machine with oracle $X$ and index $n$ to the input $m$, 
producing the correct output with probability at least $1-\frac{1}{2^{j}}$.
\end{thm}
\begin{proof}
Immediate
\end{proof}


\section{Getting by with increasingly accurate biases}

These same results can all be realised without the need for a coin with an 
infinitely precise bias. Instead, consider a variant of the PTM which is given a 
succession of coins $\seq{c_{n}}$ where the $n$-th coin is used for the $n$-th 
toss. If the probability of $c_{n}$ coming up heads is given by the rational 
probability $p_{n}$ and $\seq{p_{n}}$ converges quickly to some arbitrary real 
$p$, then all of the above results hold with only minor modifications.

If we once again approximate $p$ using the average number of times heads comes 
up in $n$ tosses, we find that the mean of $\hat{p}$ is no longer $p$, but 
$\mu$, where
\begin{equation}
\mu
= \frac{\sum_{i=1}^{n}p_{i}}{n}
\le \frac{\sum_{i=1}^{n}p+\frac{1}{2^{i}}}{n}
< \frac{np + 1}{n}
= p + \frac{1}{n}
\label{MuAndP}
\end{equation}
By a similar argument, we find the lower bound for $\mu$, and see that
\begin{equation}
p - \frac{1}{n} < \mu < p + \frac{1}{n}
\label{MuBounds}
\end{equation}
The variance is now given by
\begin{equation}
\sigma^{2} = \sum_{i=1}^{n}\frac{p_{i}(1 - p_{i})}{n^{2}} \le \frac{1}{4n}
\label{Variance}
\end{equation}

We can now once again use the Chebyshev inequality to form an upper bound for 
the probability of error. If we set $n = 2^{j+2k}$ (which is 4 times higher than 
the value of $n$ used previously), we see 
\begin{eqnarray}
P\left(\abs{\hat{p}-\mu} < \frac{1}{2^{k+1}}\right) &
\ge & 1 - \frac{1}{2^{j}}\\
P\left(\abs{\hat{p}-p} < \frac{1}{2^{k+1}} + \frac{1}{2^{j+2k}}\right) &
\ge & 1 - \frac{1}{2^{j}}\nonumber\\
P\left(\abs{\hat{p}-p} < \frac{1}{2^{k}}\right) &
\ge & 1 - \frac{1}{2^{j}}\nonumber
\label{NewVarianceInequaltiy}
\end{eqnarray}

And so this new value of $n$ suffices in this case. Replacing all later 
references to $2^{j+2k-2}$ with $2^{j+2k}$ and references to $2^{j+3k-2}$ with 
$2^{j+3k}$, all the theorems follow. It is also easy to see that we could relax 
our constraint of sequence that converges quickly to a sequence that converges 
at a rate bounded by some recursive function. Then we could calculate a 
subsequence of these coins whose probabilities would converge quickly and use 
that.


\section{Getting by with randomly chosen biases}
\label{Section:RandomBiases}

Another way that we can avoid the need for a coin with an infinitely accurate 
bias is via a probability distribution of finitely accurate biases. As in the 
previous section, we use a sequence of coins $\seq{c_{n}}$ where the $n$-th coin 
is used for the $n$-th toss. This time however, the bias on each coin will be 
chosen with an independent random trial from a fixed probability distribution. 
We will see that so long as the mean of this distribution is a non-recursive 
real, access to this randomisation extends the PTM's powers. Specifically, it 
can compute the binary expansion of the mean with arbitrarily high confidence.

We first consider the case of a discrete probability distribution, where the 
probability of choosing the bias $x_{i} \in [0,1]$ is denoted by 
$P(x_{i}\textrm{ is chosen})$. To generate the value of the $n$-th coin toss, we 
must combine the process of randomly choosing a bias with the process of tossing 
a coin with that bias. Let $z$ be a random variable representing the result of 
the coin toss, equalling 1 if the coin lands heads and 0 if tails. From the 
rules of conditional probability,
\begin{eqnarray} 
P(z=1) & = & \sum_{i}P(z=1|x_{i}\textrm{ is chosen})P(x_{i}\textrm{ is 
chosen})\\
& = & \sum_{i}x_{i}P(x_{i}\textrm{ is chosen})\nonumber\\
& = & \mu_{x}\nonumber
\label{DiscreteMean}
\end{eqnarray}

The same is true if we use a continuous distribution $\rho(x)$. In this case
\begin{eqnarray} 
P(z=1) & = & \int_{0}^{1}P(z=1|x\textrm{ is chosen})\rho(x\textrm{ is 
chosen})dx\\
& = & \int_{0}^{1}x\rho(x\textrm{ is chosen})dx\nonumber\\
& = & \mu_{x}\nonumber
\label{ContinuousMean}
\end{eqnarray}

In either case $P(z=0) = 1 - \mu_{x}$. Thus, the combined process of randomly 
choosing a bias between 0 and 1 from any distribution with mean $\mu_{x}$ and 
then flipping the appropriate coin is equivalent to the process of flipping a 
single coin with bias $\mu_{x}$. From this it is clear that one can determine 
the binary expansion of $\mu_{x}$ with arbitrary confidence using the methods of 
sections \ref{Section:ApproximatingP}--\ref{Section:PAsOracle}. Indeed, all the 
results of those sections hold for this modified type of PTM without the need 
for any additional coin tosses.

In the case of discrete distributions, it is interesting to consider how 
$\mu_{x}$ could be non-recursive. Recall that for a discrete distribution, the 
mean is defined by $\sum_{i}x_{i}P(x_{i})$ and that the recursive reals are 
closed under finite sums and products. Thus, if the distribution is finite, the 
only possibilities are that at least one of the possible biases is non-recursive 
or at least one of the associated probabilities is non-recursive. For infinite 
discrete distributions there is the additional possibility of one or both of the 
sequences $\seq{x_{i}}$ and $\seq{P(x_{i})}$ being non-recursive despite all the 
individual elements being recursive.

Therefore, this method of randomly choosing a bias from a given distribution and 
then flipping a coin with that bias allows a PTM to exceed the power of a Turing 
machine without relying upon coins with infinitely precise biases.


\section{Conclusions}

Over the course of this paper, we have shown three ways to implement an abstract 
oracle by tossing biased coins. This was achieved by demonstrating a 
computational equivalence between $o$-machines and three different classes of 
PTM. These results show that it is very careless to say that randomness does not 
increase the power of the Turing machine. While this is true of fair coins and 
recursively biased coins, they form only a set of measure zero in the space of 
all possible biased coins. Indeed, if a bias is chosen completely at random
(from a uniform distribution over $[0,1]$) then with probability one, it would
be non-recursive and thus extend the powers of a Turing machine that had access
to it.

This is not only of mathematical interest, but is particularly significant in
the study of what is physically computable. There has been continued interest
over the years about whether some form of $o$-machine might be physically
realisable \cite{Copeland:2002, Ord:2002, Kieu:2003}. A simple way to go about
implementing an oracle would be to measure some quantity, such as the distance
between two particles, with finer and finer accuracy. If this distance happened
to be a non-recursive real, we could then use the methods of section
\ref{Section:BinaryExpansion} to compute the binary expansion and use this as an
oracle set. However, such methods based on measuring continuous quantities
could quickly run into fundamental limits of Quantum Mechanics, especially 
if there exists some fundamental lengthscale such as that of the Planck scale as
demanded by some theory of Quantum Gravity.

Using randomness provides an alternative that does not run afoul of these
limitations. It allows one to measure an underlying continuous quantity with a
sequence of discrete measurements that do not individually become increasingly
accurate. It is the increasing total number of measurements that provides the
accuracy, so no particular measurement needs to be more accurate than the
quantum limits.

In fact, quantum mechanics even suggests a means of simulating such biased coin
tosses. A qubit is a generic name for a quantum system that has two possible
states and when measured, is seen to take on one of these states
randomly~\cite{Nielsen:2000}. Each state of a qubit has an associated
\emph{probability amplitude}, which is a complex number that defines the
probability that the system will be found in that state. These probability
amplitudes are allowed to be arbitrary complex numbers having moduli less than
one and thus the induced probabilities, which are squares of the moduli, are
arbitrary reals. A qubit therefore seems to be a physical
implementation of an arbitrarily biased coin.

There is, however, an important dissimilarity in that while a biased coin can be
flipped as many times as one wishes, a qubit is destroyed once its state is
determined. Furthermore, by the \emph{no cloning theorem} of quantum
mechanics~\cite{Wootters:1982}, we cannot get around this destructive
measurement by making perfect copies of the qubit.

However, the technique of section \ref{Section:RandomBiases} seems to offer a
way out. If there is any method of creating qubits with biases chosen randomly
around some non-recursive mean, then this method implements a non-recursive
oracle. Since the non-recursive values this mean could take form a set of
measure one in the space of all possible biases, this appears quite plausible
and it would seem to require an independent physical principle to force all such
methods to pick out only recursive means.

If we furthermore wish to harness this non-recursive power to compute some
particular non-recursive function, we need to know more about the non-recursive
mean around which our biases are generated. For instance, we could try to create
a PTM for deciding whether a given formula of the predicate calculus is a
tautology by using a mean that codes the set of halting Turing machines, or even
by using the {\em halting probability} $\Omega$, described by Chaitin
\cite{Chaitin:1975}, in the setting up of a qubit~\cite{Nielsen:1997}.

However, it appears to be very difficult to generate biased qubits around such a known
mean. Consider some controllable variable $\lambda$ (such
as the amount of time an electron is exposed to a magnetic field) involved in creating the
probability amplitude $a(\lambda)$ for a qubit  state and let us
suppose that we could generate this controllable variable in some distribution
$P_\lambda$ with the appropriate mean. Even then, we would still have further 
difficulties to overcome as the relationship between the bias of a qubit, 
represented by $|a(\lambda)|^2$,
and the controllable variables that determine it is inevitably non-linear. 
It is then not sufficient to control the mean of the controlled variables $\lambda$: we must
also precisely know the details both of their distributions $P_\lambda$ and of
the functions $a(\lambda)$, which can and will be
affected by generally uncontrollable quantum decoherence, to obtain the mean
of the quantum probabilities through their distributions
$P_a$,
\begin{eqnarray}
P_a &=& P_\lambda\left/\left|\frac{d|a|^2}{d\lambda}\right|\right.,
\end{eqnarray}
and it seems quite unlikely that all of these would be possible.

The use of biased coins to compute more than the Turing machine is certainly of
some physical interest and, although it is not yet clear how it could be physically
harnessed to increase our computational abilities, the close connections with
quantum theory suggest a potential for further study.

\bibliography{computation,hypercomputation,ait}

\begin{thebibliography}{10}

\bibitem{Chaitin:1975}
Gregory~J. Chaitin.
\newblock A theory of program size formally identical to information theory.
\newblock {\em Journal of the ACM}, 22(3):329--340, July 1975.

\bibitem{Copeland:2002}
B.~Jack Copeland.
\newblock Hypercomputation.
\newblock {\em Minds and Machines}, 12:461--502, 2002.

\bibitem{Gill:1977}
J.~Gill.
\newblock Computational complexity of probabilistic {Turing} machines.
\newblock {\em SIAM Journal of Computing}, 6:675--695, 1977.

\bibitem{Kieu:2003}
Tien~D. Kieu.
\newblock Computing the noncomputable.
\newblock {\em Contemporary Physics}, 44:51--71, 2003.

\bibitem{Nielsen:1997}
Michael~A. Nielsen.
\newblock Computable functions, quantum measurements, and quantum mechanics.
\newblock {\em Phys. Rev. Lett.}, 79:2915--2918, 1997.

\bibitem{Nielsen:2000}
Michael~A. Nielsen and Isaac~L. Chuang.
\newblock {\em Quantum Computation and Quantum Information}.
\newblock Cambridge University Press, Cambridge, 2000.

\bibitem{Ord:2002}
Toby Ord.
\newblock Hypercomputation: Computing more than the {Turing} machine.
\newblock Technical Report arXiv:math.LO/0209332, University of Melbourne,
  Melbourne, Australia, September 2002.
\newblock Available at http://www.arxiv.org/abs/math.LO/0209332.

\bibitem{Santos:1971}
E.~S. Santos.
\newblock Computability by probabilistic {Turing} machines.
\newblock {\em Transactions of the American Mathematical Society},
  159:165--184, 1971.

\bibitem{Turing:1936}
Alan~M. Turing.
\newblock On computable numbers, with an application to the
  entscheidungsproblem.
\newblock {\em Proceedings of the London Mathematical Society}, 42:230--265,
  1936.

\bibitem{Turing:1939}
Alan~M. Turing.
\newblock Systems of logic based on the ordinals.
\newblock {\em Proceedings of the London Mathematical Society}, 45:161--228,
  1939.

\bibitem{Wootters:1982}
W.~K. Wootters and W.~H. Zurek.
\newblock A single quantum cannot be cloned.
\newblock {\em Nature}, 299:802--803, 1982.

\end{thebibliography}
\bibliographystyle{plain}

\end{document}